\documentclass[aps, prb, 10pt, preprint, showpacs, preprintnumbers,
superscriptaddress, amsmath, amssymb]{revtex4-1}
\usepackage[pdftex]{graphicx}
\usepackage[ngerman,english]{babel}
\usepackage[ansinew]{inputenc}
\usepackage{natbib}
\usepackage{color}
\usepackage{siunitx}
\usepackage{calc}
\usepackage[colorlinks=true, pdfstartview=FitV, linkcolor=blue,
citecolor=blue, urlcolor=blue]{hyperref}

\definecolor{green2}{rgb}{.0, .58, 0}

\begin{document}


\title{Magnetic, thermal, and topographic imaging with a
  nanometer-scale SQUID-on-cantilever scanning probe}


\author{M.~Wyss} \affiliation{Department of Physics, University of
  Basel, 4056 Basel, Switzerland} \affiliation{Swiss Nanoscience Institute, University of
  Basel, 4056 Basel, Switzerland}

\author{K.~Bagani} \affiliation{Department of Physics, University of
  Basel, 4056 Basel, Switzerland}
	
\author{D.~Jetter} \affiliation{Department of Physics, University of
  Basel, 4056 Basel, Switzerland}
	
\author{E.~Marchiori} \affiliation{Department of Physics, University of
  Basel, 4056 Basel, Switzerland}
	
\author{A.~Vervelaki} \affiliation{Department of Physics, University of
  Basel, 4056 Basel, Switzerland}
	
\author{B.~Gross} \affiliation{Department of Physics, University of
  Basel, 4056 Basel, Switzerland}
	
\author{J.~Ridderbos} \affiliation{Department of Physics, University
  of Basel, 4056 Basel, Switzerland}

\author{S.~Gliga} \affiliation{Swiss Light Source, Paul Scherrer
  Institute, 5232 Villingen, Switzerland}

\author{C.~Sch\"onenberger} \affiliation{Department of Physics,
  University of Basel, 4056 Basel, Switzerland} \affiliation{Swiss
  Nanoscience Institute, University of Basel, 4056 Basel, Switzerland}

\author{M.~Poggio} \affiliation{Department of Physics, University of
  Basel, 4056 Basel, Switzerland}\affiliation{Swiss Nanoscience Institute, University of
  Basel, 4056 Basel, Switzerland} \email{martino.poggio@unibas.ch}

\date{\today}

\begin{abstract} Scanning superconducting quantum interference device
  (SQUID) microscopy is a magnetic imaging technique combining high
  field sensitivity with nanometer-scale spatial resolution.
  State-of-the-art SQUID-on-tip
  probes~\cite{finkler_self-aligned_2010,finkler_scanning_2012} are
  now playing an important role in mapping correlation phenomena, such
  as superconductivity~\cite{uri_mapping_2020} and
  magnetism~\cite{tschirhart_imaging_2021}, which have recently been
  observed in two-dimensional van der Waals materials.  Here, we
  demonstrate a scanning probe that combines the magnetic and thermal
  imaging provided by an on-tip SQUID with the tip-sample distance
  control and topographic contrast of a non-contact atomic force
  microscope (AFM).  We pattern the nanometer-scale SQUID, including
  its weak-link Josephson junctions, via focused ion beam milling at
  the apex of a cantilever coated with Nb, yielding a sensor with an
  effective diameter of 365~nm, field sensitivity of
  9.5~$\text{nT}/\sqrt{\text{Hz}}$ and thermal sensitivity of
  620~$\text{nK}/\sqrt{\text{Hz}}$, operating in magnetic fields up to
  1.0~T.  The resulting SQUID-on-lever is a robust AFM-like scanning
  probe that expands the reach of sensitive nanometer-scale magnetic
  and thermal imaging beyond what is currently possible.
\end{abstract}

\maketitle

Nanometer-scale magnetic imaging techniques, based on optical,
electron, x-ray, or scanning probe sensors, reveal magnetization
patterns, spin configurations, and current distributions.  They
provide local information about length-scales, inhomogeneity, and
interactions, that is inaccessible in bulk measurements of transport,
magnetization, susceptibility, or heat capacity.  In recent years,
these techniques have shed light on nanometer-scale phenomena such as
domain walls, magnetic and superconducting vortices, and magnetic
skyrmions.  The techniques combining the highest sensitivity and
highest spatial resolution are being used to map recently discovered
correlated states hosted in some two-dimensional (2D) van der Waals
materials and their heterostructures, because they can provide
information on quantum phases, including on the spatial variation of
order parameters, the presence of domains, and the role of defects.
Correlated states in 2D materials are extremely sensitive to disorder
and inhomogeneity.  In such a fragile environment, local measurements
-- with sensors whose characteristic size is smaller than the length
scale of the disorder -- are essential for making sense of the system.

Foremost among sensitive and high-resolution magnetic imaging
techniques is scanning superconducting quantum interference device
(SQUID) microscopy with so-called SQUID-on-tip
probes~\cite{finkler_self-aligned_2010,finkler_scanning_2012}.  These
probes consist of a SQUID fabricated by shadow evaporation or
directional sputtering of a metallic superconductor directly on the
end of a pulled quartz tip.  The resulting sensors can have diameters
as small as 50~nm and sensitivities down to
5~$\text{nT}/\sqrt{\text{Hz}}$~\cite{vasyukov_scanning_2013}.
Scanning SQUID-on-tip probes have been used to image superconducting
vortices~\cite{embon_probing_2015,ceccarelli_imaging_2019},
superparamagnetism at the
nanometer-scale~\cite{lachman_visualization_2015,anahory_emergent_2016},
and magnetic reversal in nanomagnetic
structures~\cite{vasyukov_imaging_2018,wyss_stray-field_2019}.  In 2D
systems, they have been used to visualize quantum Hall edge currents
in graphene~\cite{uri_nanoscale_2020} as well as twist-angle
disorder~\cite{uri_mapping_2020} and orbital
magnetism~\cite{tschirhart_imaging_2021} in twisted bilayer graphene.
They have also been used for thermal sensing, achieving sensitivities
better than
0.5~$\mu\text{K}/\sqrt{\text{Hz}}$~\cite{halbertal_nanoscale_2016}.
Such sensitive thermal microscopy, which was recently used to
visualize dissipation in the quantum Hall state in
graphene~\cite{halbertal_imaging_2017}, is possible only because of
its on-tip geometry, allowing the SQUID sensor to be in close
proximity to the sample, rather than a pick-up loop, inductively
coupled to a distant on-chip SQUID.  However, SQUID-on-tip probes,
which are typically coupled to stiff tuning fork resonators for
stabilization of the tip-sample distance, are only sensitive to
topography within tens of nanometers of the surface.  Their intrinsic
stiffness makes stabilization and topographic imaging challenging and
results in a high probability of the tip crashing.

Here, we use focused ion beam (FIB) milling to fabricate a
nanometer-scale SQUID at the apex of a cantilever force sensor,
combining the capabilities of a SQUID-on-tip with those of non-contact
atomic force microscopy (AFM).  All components of the SQUID, including
superconducting leads, loop, and weak-link Josephson junctions are
patterned via direct FIB milling of a sputtered Nb film, overcoming
the challenge of patterning directly on a high-aspect-ratio tip.  The
result is a SQUID-on-lever scanning probe that provides exquisite
magnetic, thermal, and topographic contrast.  The non-contact AFM
cantilever platform provides excellent control of the tip-sample
spacing, as well as topographic contrast further than 1~$\mu$m.  In
addition, FIB patterning offers flexibility in the possible designs of
the SQUID.  In contrast, the minimum size and complexity of
SQUID-on-tip probes are restricted by the fabrication process, e.g.\
not allowing for the integration of field coils for susceptibility
measurements or flux feedback.

\begin{figure}[t!]
  \includegraphics[width=0.9\textwidth]{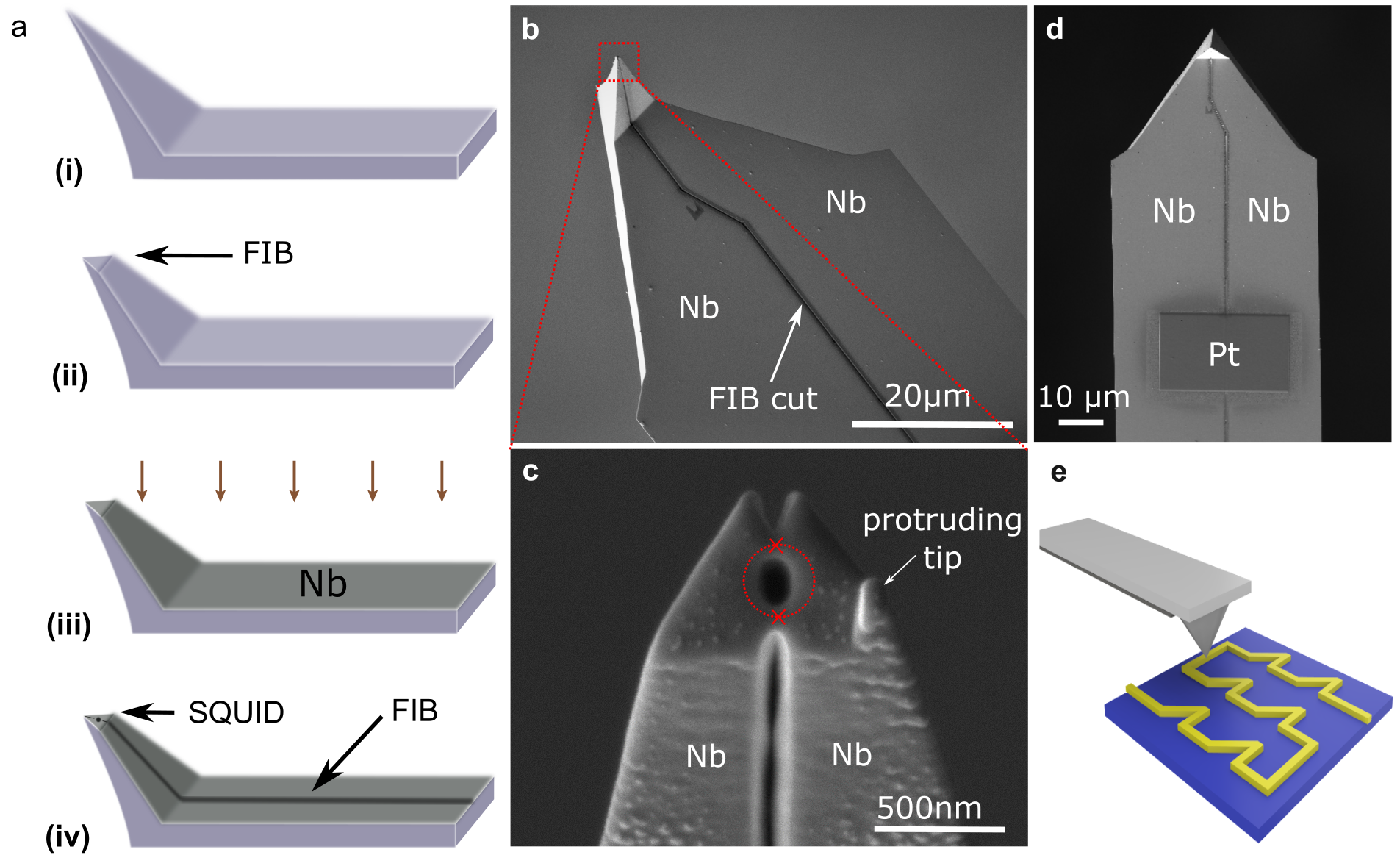}%
  \caption{SQUID-on-lever fabrication.  (a) Schematic of the steps
    for patterning a Nb SQUID-on-lever. (i) The tip of an AFM
    cantilever (ii) is milled away to create a triangular
    plateau. (iii) A Nb film is deposited on the front side of the
    cantilever and (iv) is then milled via FIB to define a SQUID.  (b)
    SEM image from below the Si cantilever showing the Nb
    film and the FIB-milled trench separating the two leads to the
    on-tip device.  (c) Zoomed-in SEM image showing the FIB-milled
    trenches and central hole defining the SQUID device.  (d) Pt shunt
    resistance connecting two Nb leads.  (e) Schematic of the
    scanning probe and sample.}%
\label{Fig1}
\end{figure}

The fabrication of a SQUID at the apex of a commercially available
Si-cantilever designed for non-contact AFM is presented schematically
in Fig.~\ref{Fig1}~(a). We use a Ga$^+$ FIB to mill away the tip of the
cantilever, leaving a triangular plateau, on which the SQUID is to be
patterned.  As part of this process, a small protrusion is left on the
side of the plateau, as seen on the right side of the scanning
electron microscope (SEM) image in Fig.~\ref{Fig1}~(c).  The role of this
feature is both to act as a sharp tip to optimize spatial resolution
in AFM mode and to prevent the SQUID from touching the sample surface
during scanning.  After this first milling step, we deposit a thin
film of Nb on the front side of the cantilever.  We then pattern the
SQUID on the coated cantilever via direct milling of the Nb film with
a Ga$^+$ FIB.  This process entails first cutting a long trench in the
film through the middle of the cantilever, from its base up to its
triangular plateau.  The SEM image in Fig.~\ref{Fig1}~(b) shows the FIB-cut
trench which defines two electrical contacts running from the base of
the cantilever to its apex.  We then mill a hole in the center of the
triangular plateau followed by a second, shorter trench.  Both the
long and short trenches are separated by the central hole and together
define two narrow constrictions forming Dayem bridge Josephson
junctions (JJs) and hence a Nb SQUID at the tip of the cantilever.
Fig.~\ref{Fig1}~(c) shows the SQUID with loop diameter of 300~nm and a
protruding tip of height 175~nm.  Finally, a thin film of Pt is
patterned via FIB-induced deposition to serve as a shunt resistor
between the two Nb leads, as shown in Fig.~\ref{Fig1}~(d).

\begin{figure}[t!]
  \includegraphics[width=0.9\textwidth]{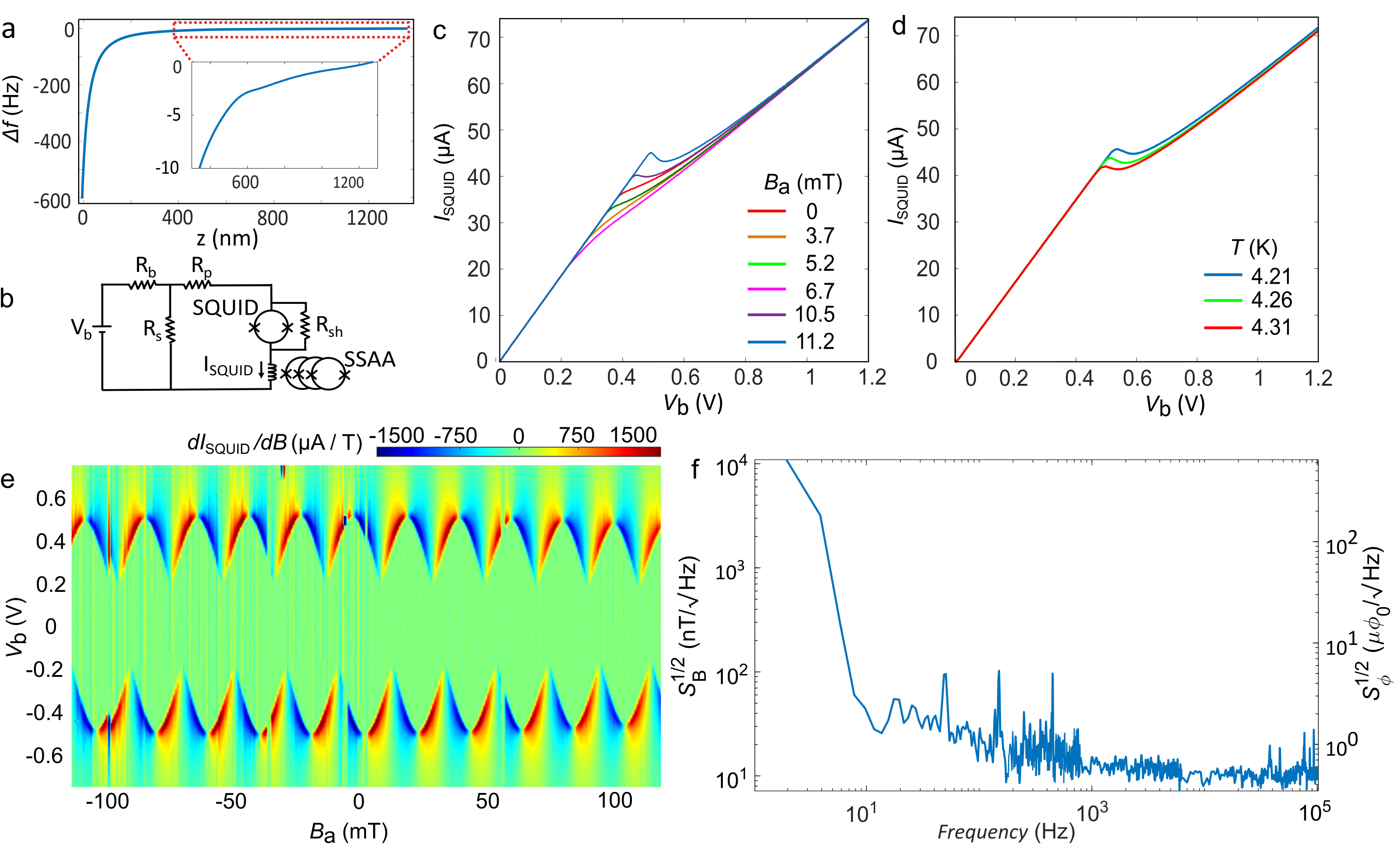}%
  \caption{SQUID-on-lever characterization. (a) $\Delta f (z)$ curve
    showing an approach of the cantilever to the sample surface.  (b)
    Schematic of the measurement circuit including inductive coupling
    to a SSAA, where $R_b = 6.1 k\Omega$, $R_s = 3\Omega$,
    $R_p = 2\Omega$, $R_{sh} = 4\Omega$.  (c) $I_{\text{SQUID}} (V_b)$
    curves of the Nb SQUID-on-lever measured at 4.2~K and at different
    $B_a$ applied along the axis of the SQUID. (d)
    $I_{\text{SQUID}} (V_b)$ curves measured at different $T$ and
    $B_a = 0$.  (e) Magnetic response function
    $d I_{\text{SQUID}}/d B$ of the SQUID as a function of both $V_b$
    and $B_a$, showing the quantum interference pattern of the
    SQUID-on-lever.  (f) Power spectral density of the device at
    4.2~K, expressed both in terms of magnetic field (left axis) and
    flux quanta (right axis).  White noise of the device reaches
    0.48~$\mu \Phi_0/\sqrt{\text{Hz}}$ or
    9.5~$\text{nT}/\sqrt{\text{Hz}}$}%
\label{Fig2}
\end{figure}

The resulting SQUID-on-lever is mounted in a custom-built scanning
probe microscope operating under vacuum at 4.2~K.  Piezoelectric
positioners move the sample under the cantilever with nanometer-scale
precision and a specialized cantilever holder allows for both
electrical contact to the SQUID-on-lever as well as mechanical
excitation.  The cantilever's flexural motion is driven by a
piezoelectric disc mounted next to the cantilever at its fundamental
resonance frequency $f_0$ of around 300~kHz and is detected using a
fiber-optic interferometer.  As the cantilever is approached to the
substrate, $f_0$ shifts, due to tip-sample interaction.  This shift
$\Delta f$ is plotted as a function of tip-sample distance $z$ in
Fig.~\ref{Fig2}~(a).  The inset of Fig.~\ref{Fig2}~(a) shows a significant response in
$\Delta f$ even for $z > 1$~$\mu$m.  As in standard non-contact AFM,
this interaction enables a controlled approach to the sample, as well
as the ability to safely scan within a few tens of nanometers of the
sample surface by feeding back and stabilizing at a constant
$\Delta f$.

We characterize the magnetic and thermal sensitivity of the
SQUID-on-lever at 4.2~K using the electrical circuit in Fig.~\ref{Fig2}~(b).
We operate the SQUID in unconventional voltage-bias mode with a large
bias resistor, $R_b$ and a small shunt resistor, $R_s$.  The circuit
includes a parasitic resistance $R_p$ in series with the SQUID due to
the wires and contacts.  The Pt shunt resistance $R_{sh}$ is connected
in parallel with the SQUID and helps to reduce hysteresis in the
SQUID's $I_{\text{SQUID}}(V_b)$ response, by damping its high
frequency dynamics.  We measure the current through the SQUID
$I_{\text{SQUID}}$ using a cold SQUID series array amplifier (SSAA)
working in feedback mode.  Figs.~2~(c) and (d) show plots of
$I_{\text{SQUID}}(V_b)$ measured at different applied magnetic fields
and temperatures respectively.  The measurements show a maximum
critical current $I_c = 45$~$\mu$A at 4.2~K.  A magnetic field $B_a$
applied along the axis of the SQUID loop $\hat{z}$ produces a
modulation of the critical current of 25~$\mu$A, which persists up to
1.0~T.  From measurements of $I_{\text{SQUID}}(V_b)$ at different
temperatures $T$, shown in Fig.~\ref{Fig2}~(d), $I_c$ is seen to decrease with
increasing $T$.

In Fig.~\ref{Fig2}~(e), we plot the magnetic response function
$d I_{\text{SQUID}}/d B$, where $B$ is the total magnetic field, as a
function of $V_b$ and $B_a$, showing a pronounced quantum interference
pattern with a period of 20~mT.  This period corresponds to an
effective SQUID loop diameter of 365~nm, which is slightly larger than
the diameter measured in the SEM image in Fig.~\ref{Fig1}~(c).  This mismatch
is likely due to both magnetic field lensing by the superconductor on
the triangular plateau and the suppression of superconductivity caused
by ion implantation in the Nb near the milled regions.  The
interference pattern shows a large asymmetry between positive and
negative $V_b$, due to the asymmetry of the two JJs.  By appropriate
choice of $V_b$, this asymmetry increases the range of applied fields,
in which the SQUID is sensitive, including at $B_a = 0$.

Fig.~\ref{Fig2}~(f) shows the spectral density of the SQUID's noise both in
terms of magnetic flux and magnetic field measured at $B_a = 0 $ and
$V_b = 0.5$~V.  The spectrum is dominated by $1/f$-like noise at low
frequencies and white noise $S^{1/2}_\Phi $ under
1~$\mu \Phi_0/\sqrt{\text{Hz}}$ above a few hundred Hz.  At 10~kHz and
in the sensitive regions of the magnetic response function, the
magnetic field sensitivity reaches $S^{1/2}_B $
9.5~$\text{nT}/\sqrt{\text{Hz}}$, which corresponds to a flux
sensitivity $S^{1/2}_\Phi $ of 0.48~$\mu \Phi_0/\sqrt{\text{Hz}}$.
This sensitivity is significantly better than previously reported
SQUID-on-tip sensors made from Al~\cite{finkler_self-aligned_2010},
Nb~\cite{finkler_scanning_2012}, and
Mo$_{66}$Re$_{34}$~\cite{bagani_sputtered_2019}, but does not attain
the sensitivity of state-of-the-art Pb
SQUID-on-tips~\cite{vasyukov_scanning_2013}. 

We determine the thermal sensitivity of the device, by analyzing
measurements of $I_{\text{SQUID}}$($V_b$) taken at different $T$, as
in Fig.~\ref{Fig2}~(d).  We calculate the thermal response
$d I_{\text{SQUID}}/d T $ from the slope of $I_{\text{SQUID}}(T)$
curve at a fixed bias voltage $V_b$.  For $V_b = 0.55$~V and at
$B_a = 0$ and $T = 4.2$~K, $d I_{\text{SQUID}}/d T = -24.2$~$\mu A/K$.
This response and the corresponding white noise floor of the device
yields a thermal sensitivity better than 620~nK$/\sqrt{\text{Hz}}$,
which is comparable to the the state-of-the-art Pb SQUID-on-tip
sensors.  

\begin{figure}[t!]
  \includegraphics[width=0.9\textwidth]{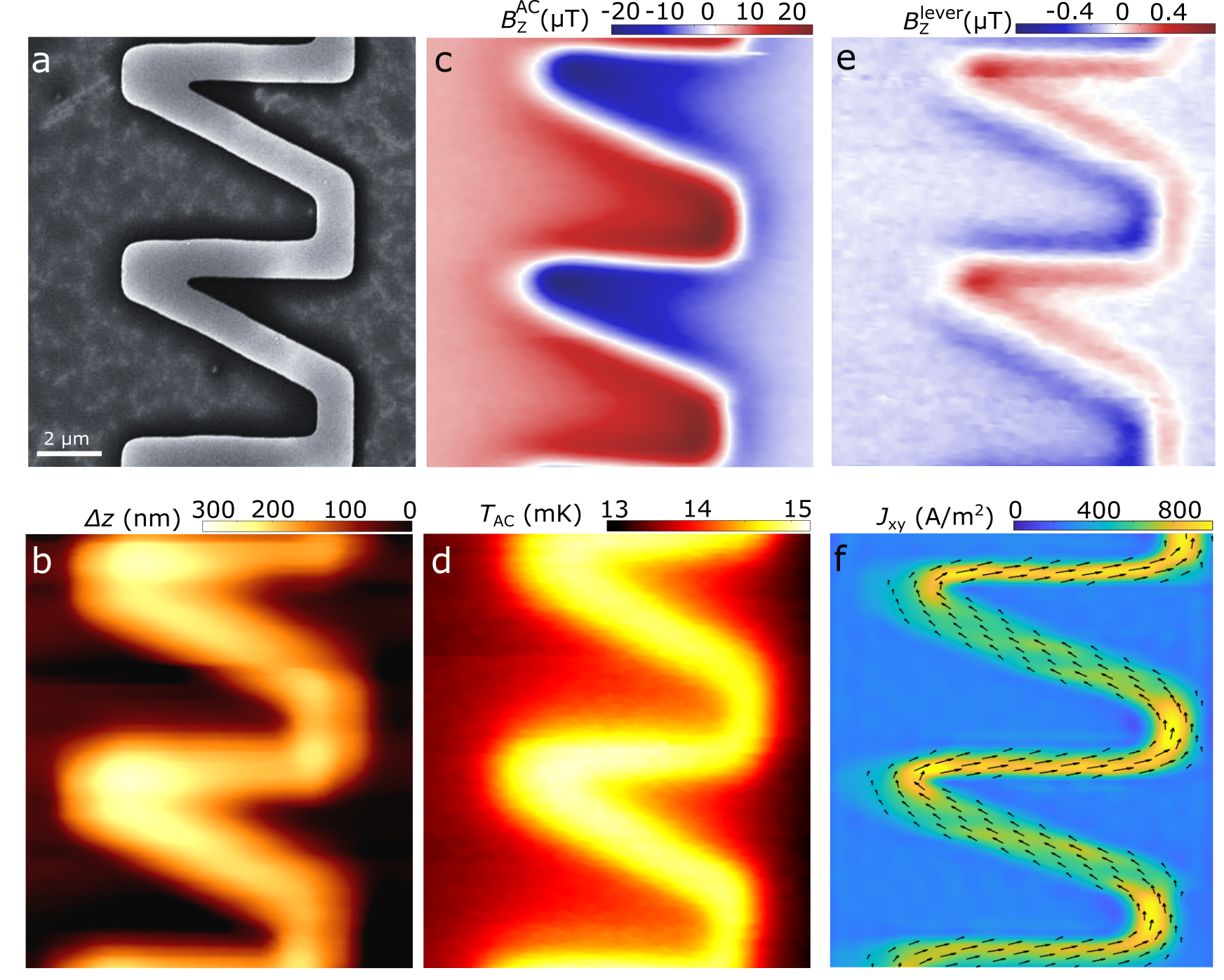}%
  \caption{Scanning probe images of a current-carrying wire taken by a
    SQUID-on-lever.  (a) SEM image of the 750-nm-wide and 300-nm-thick
    saw-tooth-shaped Au wire patterned on a Si substrate.  (b)
    Non-contact AFM image taken using the SQUID-on-lever in
    constant-frequency mode showing the sample topography $\Delta z$
    taken at a tip-sample spacing $z = 140$~nm.  Scanning probe images
    made by the SQUID-on-lever at constant height from the substrate
    corresponding to $z = 345$~nm of (c) the $z$-component of the
    Biot-Savart field $B_z^{\text{AC}}$ and (d) the AC temperature
    modulation $T_{\text{AC}}$ produced by
    $I_{\text{AC}} = 100$~$\mu$A of current flowing through the wire.
    (e) Image of $z$-component of the magnetic field
    $B_z^{\text{lever}}$ at the cantilever oscillation frequency
    produced by $I_{\text{DC}} = 200$~$\mu$A of DC.  This signal is
    proportional to the spatial derivative of $B_z$ along the
    cantilever oscillation direction $\hat{z}$.  (f) A map of the
    current density $\mathbf{J}_{xy}$ flowing through the wire
    calculated from the measurement of $B_z^{\text{AC}}$ in (c).  The
    color scale indicates the magnitude and the arrows the direction
    of $\mathbf{J_{xy}}$.}
\label{Fig3}
\end{figure}

\begin{figure}[t!]
  \includegraphics[width=0.9\textwidth]{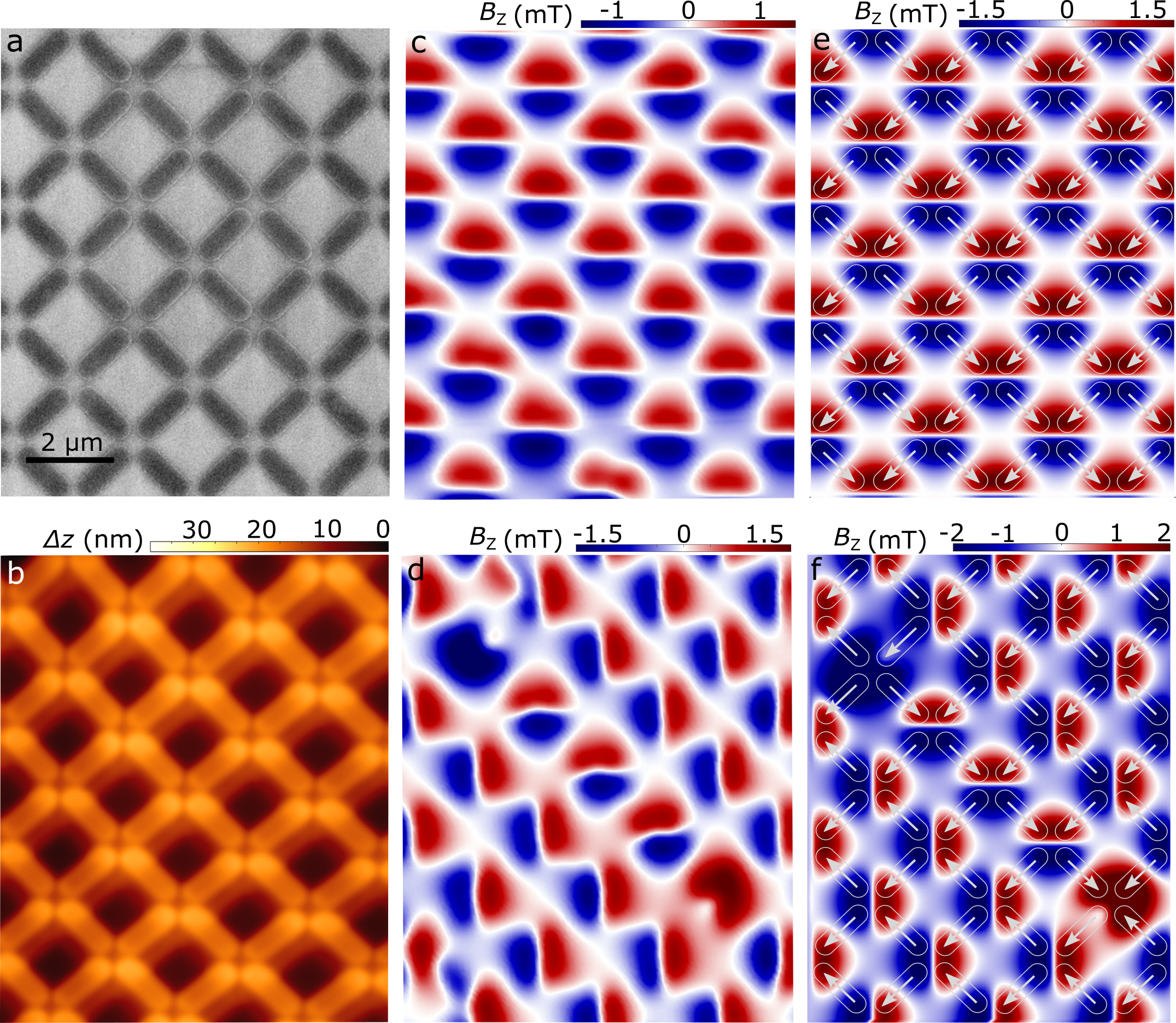}%
  \caption{Scanning probe images of artificial spin ice taken by a
    SQUID-on-lever.  (a) SEM image of the nanomagnet array consisting
    of a square lattice of 20-nm-thick Py islands with a lateral size
    of $550 \text{ nm} \times 1550 \text{ nm}$ patterned on a Si
    substrate.  (b) Non-contact AFM image taken in constant-frequency
    mode showing the sample topography $\Delta z$ at a tip-sample
    spacing $z = 65$~nm. (c) Magnetic stray field along the
    $z$-direction $B_z$ measured by the scanning SQUID-on-lever at
    constant height corresponding to $z = 275$~nm from the substrate.
    The spin ice is in remanence after saturation along $\hat{y}$ and
    (d) after the execution of a minor hysteresis loop, showing the
    signatures of separated magnetic charges and their corresponding
    numerical simulations in (e) and (f).  The position of the
    nanomagnets is shown schematically and arrows show the orientation
    of their magnetic moment.}%
\label{Fig4}
\end{figure}

\noindent In order to demonstrate the capabilities of the
SQUID-on-lever as a scanning probe, we first image a current-carrying
Au-wire patterned on a Si substrate.  We image the topography of the
wire with our SQUID-on-lever by implementing a standard non-contact
AFM technique: scanning in the $xy$-plane while feeding back on the
piezoelectric scanner controlling the the tip-sample spacing $z$ with
a correction $\Delta z$ to maintain a constant frequency shift
$\Delta f$ (constant frequency mode). In Fig.~\ref{Fig3}~(b) we present the AFM
image of the Au-wire scanned at a constant $\Delta f = -5 Hz$ which
corresponds to a tip-sample spacing of about $z = 140$~nm.
The low spatial resolution of the image is set by the small $\Delta f$
set-point resulting in a large tip-sample spacing and wide dimension
of the triangular plateau (a higher resolution topographic image is
presented in Fig.~\ref{Fig4}~(a)).  By measuring the SQUID
response 
while scanning over the same area at a fixed height corresponding to
$z = 345$~nm over the substrate, we image the Biot-Savart field and
the thermal dissipation produced by alternating current (AC) flowing
through the wire.  Note that the SQUID-to-sample spacing
$z_{\text{SQUID}} = z + 175$~nm, due to the height of the protruding
tip.  Fig.~\ref{Fig3}~(c) shows the $z$-component of the Biot-Savart field
$B_z^{\text{AC}}$ produced by $I_{\text{AC}} = 100$~$\mu$A of current
at $f_{\text{AC}}= 4.17$~kHz.  Sensitivity to magnetic field is
achieved by choosing a combination of $B_a$ and $V_b$, in which the
SQUID-on-lever has a strong magnetic response, which is plotted in
Fig.~\ref{Fig2}~(d), and measuring the first harmonic response of the SQUID at
$f_{\text{AC}}$.  By introducing a few mbar of $^4$He exchange-gas
into the vacuum chamber, we also make our SQUID-on-lever sensitive to
the local temperature modulation $T_{\text{AC}}$ produced by Joule
heating as AC flows through the Au wire.  The ability to sensitively
map $T_{\text{AC}}$ is made possible by the thermal coupling of the
sensor to the sample.  In particular, this exchange-gas-mediated
coupling must be stronger than the coupling of the sensor to its
support.  To be a non-invasive probe, this coupling must also be
weaker than the coupling of the sample with its substrate.  Both of
these conditions are met in our system, where the sensor is very
weakly coupled to its support, because of the small geometrical
cross-section of the cantilever and the absence of thermal
conductivity along the superconducting film.  Because
$T_{\text{AC}} \ll T$, where the bath temperature $T = 4.2$~K,
$T_{\text{AC}}$ is in the small signal limit and is proportional to
the power dissipated, i.e.\ the square of the flowing
current~\cite{halbertal_nanoscale_2016}.  Therefore to image
$T_{\text{AC}}$, we map the second harmonic response of the SQUID to
the current at $2 f_{\text{AC}}$~\cite{halbertal_nanoscale_2016}, as
shown in Fig.~\ref{Fig3}~(d) with $I_{\text{AC}} = 100$~$\mu$.

\noindent Fig.~\ref{Fig3}~(e) shows another imaging mode available to the
SQUID-on-lever probe: we actuate the cantilever's fundamental mode at
$f_0 = 282$~kHz with an amplitude of 15~nm and measure the
$z$-component of the magnetic field at this frequency
$B_z^{\text{lever}}$.  The resulting image, measured at a constant
height corresponding to $z = 345$~nm over the substrate, is
proportional to the spatial derivative of the magnetic field along the
cantilever oscillation direction $d B_z / d z$ produced by
$I_{\text{DC}} = 200$~$\mu$A of direct current (DC) flowing through
the wire. The use of lock-in techniques to demodulate and spectrally
filter the resulting signal substantially reduces $1/f$-noise, which
dominates DC measurements of $B_z$. Imaging such spatial magnetic
field derivatives also increases the maximum sensitivity towards
smaller features sizes, compared with imaging magnetic
fields~\cite{marchiori_technical_2021}.  In fact, detection of
magnetic spatial derivatives is the standard mode for MFM and has also
been implemented using a tuning fork resonator in scanning
SQUID-on-tip microscopy~\cite{uri_nanoscale_2020}, in order to
increase sensitivity to small DC features in $B_z$ or local $T$.

\noindent The image of the Biot-Savart field produced by current flowing through
the Au wire, shown in Fig.~\ref{Fig3}~(c), can be reconstructed into a
map of the current density $\mathbf{J}_{xy}$ by inverting the
Biot-Savart law~\cite{roth_using_1989,chang_nanoscale_2017}.  Although
3D current densities do not produce a unique magnetic stray field
patterns and can therefore not be determined by stray magnetic field
imaging alone, in-plane current densities or current densities which
are uniform throughout the thickness do.  By assuming a uniform
$\mathbf{J}_{xy}$ throughout the thickness of the Au wire, we
reconstruct the current density shown in Fig.~\ref{Fig3}~(f).

In order to demonstrate the probe's applicability to nanomagnetic
samples, we image an artificial spin
ice~\cite{wang_artificial_2006,skjaervo_advances_2020}, shown in
Fig.~\ref{Fig4}~(a), consisting of a lattice of lithographically patterned
nanometer-scale magnets, geometrically arranged to exhibit magnetic
frustration.  These samples are characterized by collective
excitations such as emergent magnetic
monopoles~\cite{wang_artificial_2006,mengotti_real-space_2011,farhan_emergent_2019}
or spin wave
modes~\cite{iacocca_reconfigurable_2016,iacocca_topologically_2017}.
In Fig.~\ref{Fig4}~(b), we show the sample's topography imaged using the
SQUID-on-lever as a non-contact AFM in constant-frequency mode. We use
a set-point of $\Delta f = -120$~Hz, corresponding to $z = 65$~nm.
The resulting image shows the array of nanomagnets with a resolution
on the order of 50~nm.  The brightest and highest resolution features
in the image are due to the protruding tip patterned on the
SQUID-on-lever, shown in Fig.~\ref{Fig1}~(c).  Fainter and lower resolution
`shadows' correspond to the sample's interaction with the larger and
more distant plateau section of the tip, on which the SQUID is
patterned.

Fig.~\ref{Fig4}~(c) shows a map taken at a constant height corresponding to
$z = 275$~nm above the substrate ($z_{\text{SQUID}}=450$~nm) over the
same region the sample.  We measure the SQUID response
$I_{\text{SQUID}}$ to the magnetic stray field $B_z$ produced by the
spin ice at remanence following saturation by an applied magnetic
field along $\hat{y}$, $B_{ay} = 100$~mT.  This corresponds to a
two-in/two-out type-II state of the square ice.  The measurement is in
good agreement with the expected $B_z(x,y)$ obtained from
corresponding micromagnetic simulations~\cite{kakay_speedup_2010}
which is presented in Fig.~\ref{Fig4}~(e).  The position and magnetization of
the simulated islands is overlaid onto this map, in order to clarify
the origin of the stray field pattern.  Following a minor hysteresis
loop (applying a magnetic field $B_{ax} = 9.3$~mT and then reducing to
zero), we record another image of the same region, shown in
Fig.~\ref{Fig4}~(d).  Whereas magnetic charges are compensated at each
four-nanomagnet vertex in the type-II configuration (Figs.~4~(c) and
(e)), after the minor loop, two particular vertices are left with
uncompensated charges (Fig.~\ref{Fig4}~(d)).  A string of vertices with
compensated charges, whose distribution is rotated with respect to the
rest of the array, connect these poles.  Fig.~\ref{Fig4}~(f) shows the results
of corresponding micromagnetic simulations matching the measurements.
The orientation of each nanomagnet's magnetic moment, reveals that the
poles correspond to `monopoles' or type-III vertices.  The asymmetric
distribution of the stray fields around these points is due to the
three-in/one-out or one-in/three-out configuration of the vertices.

Previous attempts at realizing on-tip SQUIDs have all involved either
micrometer-sized sensors and/or coupling to stiff tuning fork
resonators, which preclude topographic contrast or tip-sample
stabilization for spacings larger than tens of
nanometers~\cite{finkler_scanning_2012,shibata_imaging_2015,kirtley_scanning_2016}.
The principle obstacle to realizing a nanometer-scale SQUID-on-lever
has been the difficulty of patterning a tiny superconducting circuit
on a high-aspect-ratio tip.  Our work shows that FIB patterning --
with its ability to mill superconducting materials on the
nanometer-scale and on non-planar surfaces -- provides a solution to
this technical challenge.  In addition, it may be possible to pattern
even smaller SQUID-on-lever devices via focused ion beam or focused
electron beam induced deposition of superconducting nanostructures,
e.g.\ from WC~\cite{cordoba_magnetic_2013}, or via the structurally
modification of superconductors via a He$^+$ FIB, e.g.\ as
demonstrated in YBCO~\cite{muller_josephson_2019}.  Moreover, more
complex SQUID devices that those demonstrated here, including
modulation lines or coils for susceptibility measurements, could also
be patterned using FIB milling.  Such increased device complexity
could significantly expand the capabilities of scanning probe
microscopy, which -- although spectacularly successful -- is typically
carried out with simple probes, such as sharp conducting, insulating,
or magnetic tips.

\section*{Methods}

\textit{SQUID-on-lever fabrication}.  The non-contact AFM cantilever
(Nanosensors ATEC-NC) used here is made from doped single-crystal Si
and has a fundamental mechanical resonance frequency $f_0 = 335$~kHz,
spring constant $k = 45$~$\text{N}/\text{m}$, length $l = 160$~$\mu$m,
widths $w = 45$~$\mu$m, and thickness $t = 4.6$~$\mu$m.  We deposit
5~nm of Ti, 50~nm of Nb, and 2~nm of Pt on the front-side of the
cantilever via sputtering deposition.  Ti and Pt are used as sticking
and protection layers, respectively.  Prior to sputtering, the chamber
is pumped to a pressure of $5 \times 10^{-10}$~mbar.  200~W of DC
power are used during sputtering with a flow of 40~sccm of Ar gas and
a pressure of 4~mbar.  After sputtering, we evaporate a 10-nm-thick
layer of Au to reduce ion implantation in subsequent FIB-milling
steps.  FIB-milling is done in a dual-beam SEM and FIB (FEI Helios
Nano Lab 650) in a pressure of $1 \times 10^{-5}$~mbar.  Milling of
the cantilever tip into a triangular plateau is done with a beam
current of 1.1~pA.  Milling of the trenches is carried our in 5 steps
with different currents optimized to protect the Nb from ion
implantation.  For the long trench, far from the device area, a beam
voltage of 30~kV and beam current of 0.4~nA is used.  Closer to the
final device, the beam current is reduced in steps down to 1.1~pA at
the plateau.  The rectangular Pt shunt resistance, shown in
Fig.~\ref{Fig1}~(d), measures $30 \times 20$-$\mu$m$^2$, is 20-nm-thick, and is
patterned via FIB-induced deposition using a voltage of 30~kV and beam
current of 80~pA.

\textit{Sample fabrication}.  The current-carrying Au wire is patterned on a Si substrate
using e-beam lithography and deposited via e-beam deposition.  The
artificial spin ice, consisting of an array of permalloy
(Ni$_{83}$Fe$_{17}$) nanomagnets in a square ice geometry, were
fabricated on a non-magnetic Si (100) substrate via electron-beam
lithography, thermal evaporation at room temperature under a pressure
of $10^{-6}$~mbar, and lift-off.  The nanomagnets are
1.55~$\mu$m-long, 560~nm-wide, and have a lattice constant of
1.5~$\mu$m.  A 2-nm-thick Al capping layer is evaporated on top of the
permalloy to prevent oxidation.  The thickness of the nanomagnets is
measured by AFM to be 10~nm on average across the entire array,
without taking the capping layer into account.  The thickness is
chosen to ensure that the nanomagnets are in a single-domain
state~\cite{wyss_stray-field_2019}.

\textit{Scanning probe microscopy}.  The scanning probe microscope
operates under high vacuum in a $^4$He cryostat and employs
piezoelectric walkers and scanners (Attocube ANPx311/LT/HV,
ANSxy100lr/LT/HV) to move the sample.  The fiber-optic interferometer
consists of a 1550-nm diode laser, a 95:5 fiber-optic coupler, a fast
and a custom-built objective focusing light onto the cantilever to an
8~$\mu$m spot.  The resulting low-finesse Fabry-Perot interferometer
acts as a sensitive sensor of the cantilever's flexural motion, where
the interference intensity is measured by a fast photo-receiver with
an effective 3-dB-bandwidth of 800~kHz.  The cavity is stabilized
against drift using a PID loop (Zurich Instruments MFLI) controlling
the laser temperature.  The incident power of around 1~$\mu$W does not
significantly heat the cantilever, as confirmed by measurements of
laser power dependence and mechanical thermal motion.  During approach
and scanning a phase-locked loop (Zurich Instruments MFLI) is used to
excite and determine cantilever's mechanical resonance frequency
$f_0$.  $f_0$.

\textit{Current reconstruction and micromagnetic simulations}.
Reconstruction of the current flowing through the Au wire is carried
out using the $B_z(x,y)$ measurements and following the procedures in
refs.~\cite{roth_using_1989,chang_nanoscale_2017}.  Micromagnetic
simulations are performed with the \textit{Mumax} software
package~\cite{Vansteenkiste2014}, which employs the
Landau-Lifshitz-Gilbert micromagnetic formalism using
finite-difference discretization.  The size, shape, and material
parameters of the simulated nanomagnets correspond to those of the
measured spin ice sample.  We use typical parameters for permalloy:
saturation magnetization $\mu_0 M_{S} = 1$~T, exchange constant
$A_{ex} = 1.3 \times 10^{-11}$~$\text{J}/\text{m}$, and no
magnetocrystalline anisotropy, $K = 0$.  A temperature $T = 0$ is
assumed.  The structure is discretized into cells of size
$9.19 \text{ nm}\times 9.19 \text{ nm}\times 10 \text{ nm}$.

\section*{Author Contributions}

M.P and M.W. conceived and designed the experiment.  M.W., D.J. and
K.B. fabricated the SQUID-on-lever.  K.B. and D.J. performed the
characterization measurements with help from M.W and E.M.  The
scanning probe measurements were carried out by K.B., M.W., D.J., and
A.V.  D.J., B.G., and S.G. performed micromagnetic simulations.
J.R. and C.S. provided the Nb sputtering system.  S.G. fabricated the
nanomagnet arrays.  M.P. wrote the manuscript with input from K.B. and
M.W.  All authors discussed the results and commented on the
manuscript.  M.P. supervised the project.  The authors declare that
they have no competing financial interests.

\begin{acknowledgments}
  We thank Roy Haller for contributions to the sputtering high-quality
  Nb and Jos\'e Mar\'ia De Teresa for important suggestions.  We also
  thank Sascha Martin and his team in the machine shop of the Physics
  Department at the University of Basel for help building the
  measurement system.  We acknowledge the European Commission under
  H2020 FET Open grant `FIBsuperProbes' (number 892427), H2020 ERC
  Advanced grant TopSupra, the SNF under Grant 200020-178863, the
  Swiss Nanoscience Institute, the Kanton Aargau, and the NCCR Quantum
  Science and Technology (QSIT).
\end{acknowledgments}

\end{document}



\title{Supplementary Information: \\ Magnetic, thermal, and topographic imaging with a
nanometer-scale SQUID-on-cantilever scanning probe}



\author{M.~Wyss} \affiliation{Department of Physics, University of
  Basel, 4056 Basel, Switzerland} \affiliation{Swiss Nanoscience Institute, University of
  Basel, 4056 Basel, Switzerland}

\author{K.~Bagani} \affiliation{Department of Physics, University of
  Basel, 4056 Basel, Switzerland}
	
\author{D.~Jetter} \affiliation{Department of Physics, University of
  Basel, 4056 Basel, Switzerland}
	
\author{E.~Marchiori} \affiliation{Department of Physics, University of
  Basel, 4056 Basel, Switzerland}
	
\author{A.~Vervelaki} \affiliation{Department of Physics, University of
  Basel, 4056 Basel, Switzerland}
	
\author{B.~Gross} \affiliation{Department of Physics, University of
  Basel, 4056 Basel, Switzerland}
	
\author{J.~Ridderbos} \affiliation{Department of Physics, University
  of Basel, 4056 Basel, Switzerland}

\author{S.~Gliga} \affiliation{Swiss Light Source, Paul Scherrer
  Institute, 5232 Villingen, Switzerland}

\author{C.~Sch\"onenberger} \affiliation{Department of Physics,
  University of Basel, 4056 Basel, Switzerland} \affiliation{Swiss
  Nanoscience Institute, University of Basel, 4056 Basel, Switzerland}

\author{M.~Poggio} \affiliation{Department of Physics, University of
  Basel, 4056 Basel, Switzerland}\affiliation{Swiss Nanoscience Institute, University of
  Basel, 4056 Basel, Switzerland} \email{martino.poggio@unibas.ch}



\date{\today}

\maketitle

\section{Modulation of SQUID critical current with field}

\begin{figure}[h]
	\includegraphics[width=0.8\textwidth]{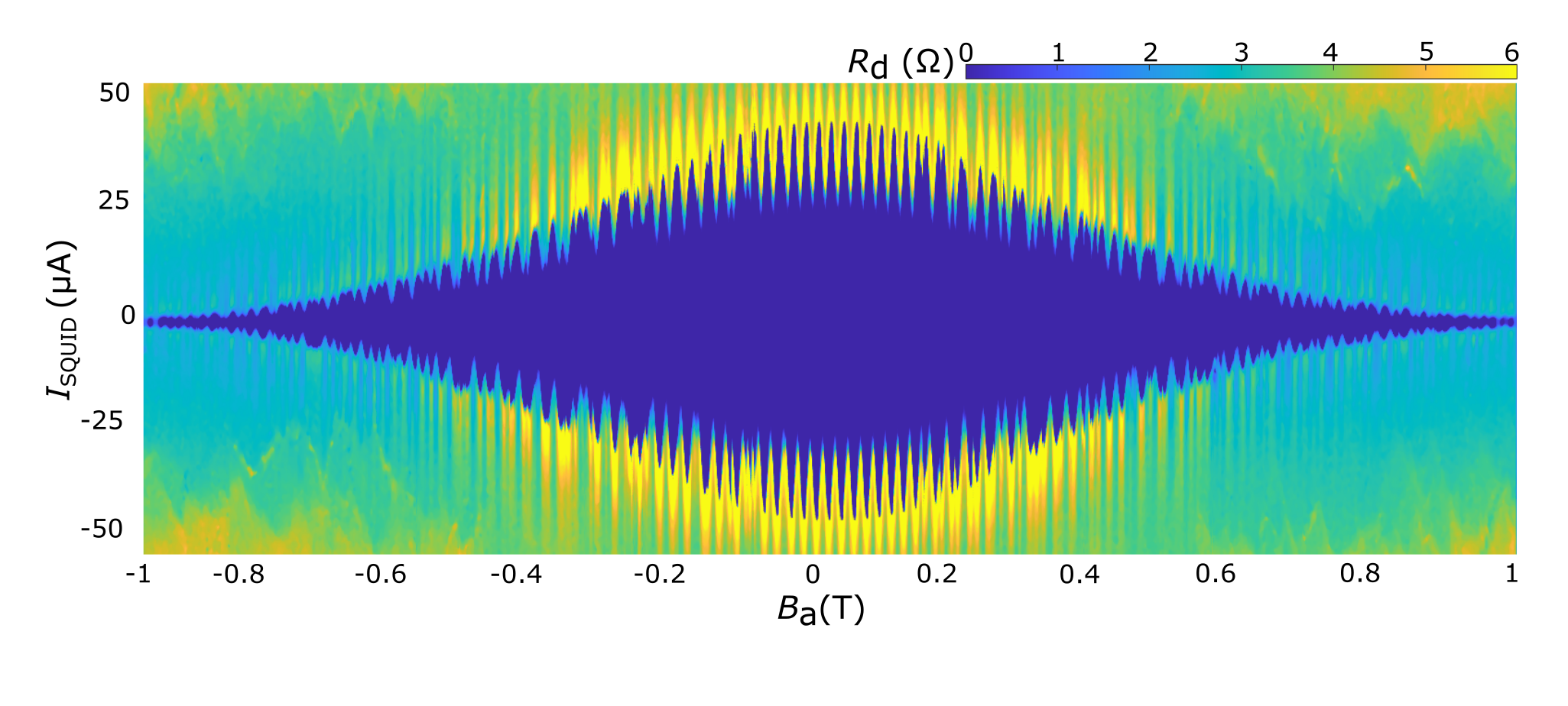}%
	\caption{Differential resistance $R_d = dV_{SQUID}/dI_{SQUID}$ as
		a function of $B_a$ and $I_{SQUID}$ of the device
		showing SQUID interference pattern up to 1T.}%
	\label{S1}
\end{figure}

Fig.~\ref{S1} shows the differential resistance of the SQUID $R_d = dV_{SQUID}/dI_{SQUID}$ as a function of $B_a$ and $I_{SQUID}$ obtained from the measured $I_{SQUID}$ vs. $V_b$ curves. The SQUID shows  quantum interference pattern up to $B_a = 1$~T. The interference pattern has a period of
20~mT, which translates to an effective SQUID loop diameter of
365~nm. The effective SQUID loop diameter is derived from
$D=2\sqrt{\Phi_0/\pi\Delta B}$, where $\Phi_0$ is the flux quantum and
$\Delta B$ is the period of the interference pattern.

\section{Field dependence of magnetic sensitivity}

\begin{figure}[h]
	\includegraphics[width=0.5\textwidth]{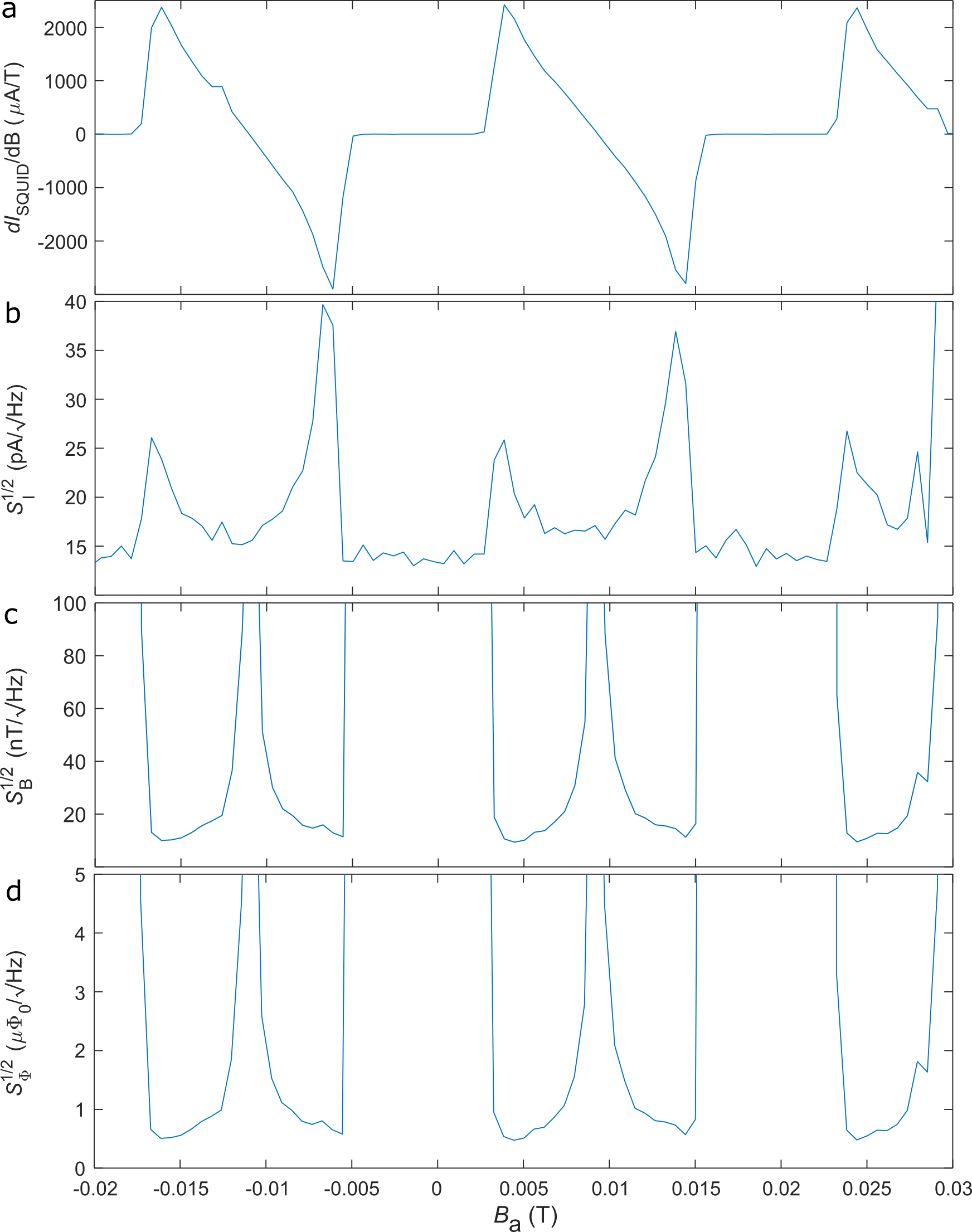}%
	\caption{ (a) and (b) Magnetic response function
          $dI_{SQUID}/dB$ and current noise $S^{1/2}_I$ at a constant
          bias voltage $V_b= 0.4$~V plotted as a function of applied
          magnetic field $B_a$. (c) and (d) Corresponding field noise
          and flux noise of the SQUID-on-lever as a function of
          $B_a$. }%
	\label{S2}
\end{figure}

The magnetic sensitivity of the device is obtained from the magnetic
response function $dI_{SQUID}/dB$ and the white noise floor of the
device. The magnetic response is derived from $I_{SQUID}(V_b)$ curve
measured at different applied magnetic fields $B_a$. Fig.~\ref{S2}~(a)
shows the plot of magnetic response with the applied magnetic field at
a fixed bias $V_b = 0.4$~V. At this constant bias voltage, the
spectral density of the current noise $S^{1/2}_I$ at 10~kHz is
measured for the same range of $B_a$, as shown in
Fig~\ref{S2}~(b). The magnetic field noise of the device is then given
by $S^{1/2}_B=S^{1/2}_I/|dI_{SQUID}/dB|$. The corresponding flux noise
is obtained form $S^{1/2}_{\Phi}=S^{1/2}_B \times (A/\Phi_0)$, where
$\Phi_0$ is the flux quantum and $A$ is the effective area of the
SQUID loop. The field noise and the flux noise of the device as a
function of $B_a$ are shown in Figs.~\ref{S2}~(c) and (d),
respectively. The flux and field noise decrease in the high magnetic
response regions of the interference pattern.  In the sensitive
regions, the device attains a field sensitivity
$S^{1/2}_B = 9.5$~$\text{nT}/\sqrt{\text{Hz}}$ or flux sensitivity
$S^{1/2}_{\Phi} = 0.48$~$\mu \Phi_0/\sqrt{\text{Hz}}$.

\section{Thermal response}

\begin{figure}[h]
  \includegraphics[width=0.4\textwidth]{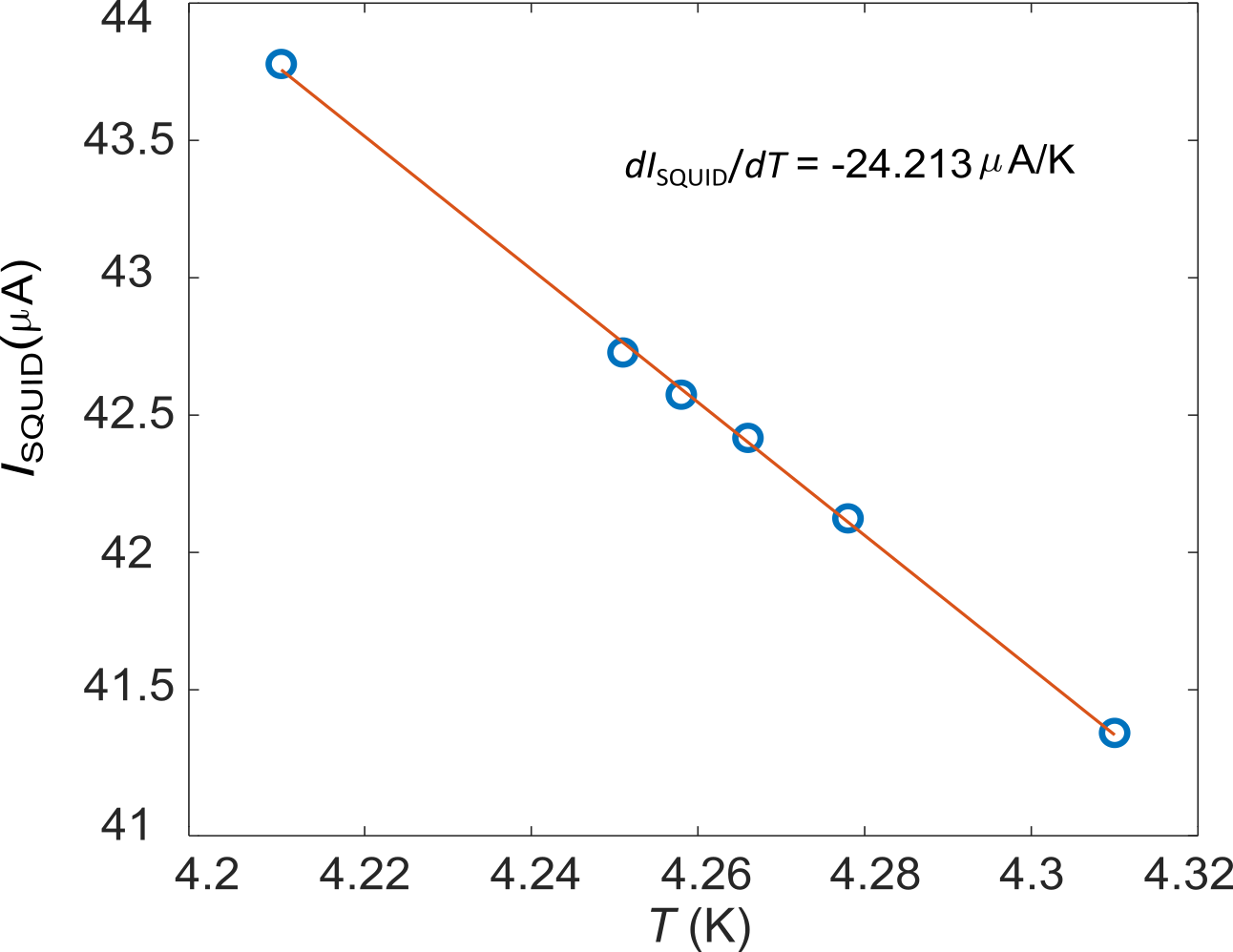}%
  \caption{ $I_{SQUID}$ vs.\ temperature $T$ at $V_b =
    0.55$~V. Thermal response of -24.21~$\text{uA}/\text{K}$ is
    obtained from the slope.}%
	\label{S3}

\end{figure}

In order to determine thermal sensitivity of the device, we fix the
voltage bias to make the SQUID-on-lever maximally sensitive to changes
in temperature.  As shown in Fig.~2~(d) of the main text, where
$I_{SQUID}(V_b)$ is measured at $B_a = 0$ and different $T$, this
occurs for $V_b = 0.55$~V.  At this bias voltage, we plot
$I_{SQUID}(T)$ in Fig.~\ref{S3}.  The slope of this curve yields the
thermal response of the device $dI_{SQUID}/dT=
-24.2$~$\mu \text{A}/\text{K}$ at 4.2~K. By measuring the white
current noise $S^{1/2}_I$, we determine the temperature noise
$S^{1/2}_T=S^{1/2}_I/|dI_{SQUID}/dT|$ of the device.  The
SQUID-on-lever reaches a thermal sensitivity
$S^{1/2}_T= 620$~$\text{nK}/\sqrt{\text{Hz}}$ at zero magnetic field
and 4.2~K.

\section{Scanning magnetic imaging of the current-carrying wire}

\begin{figure}[h]
	\includegraphics[width=0.6\textwidth]{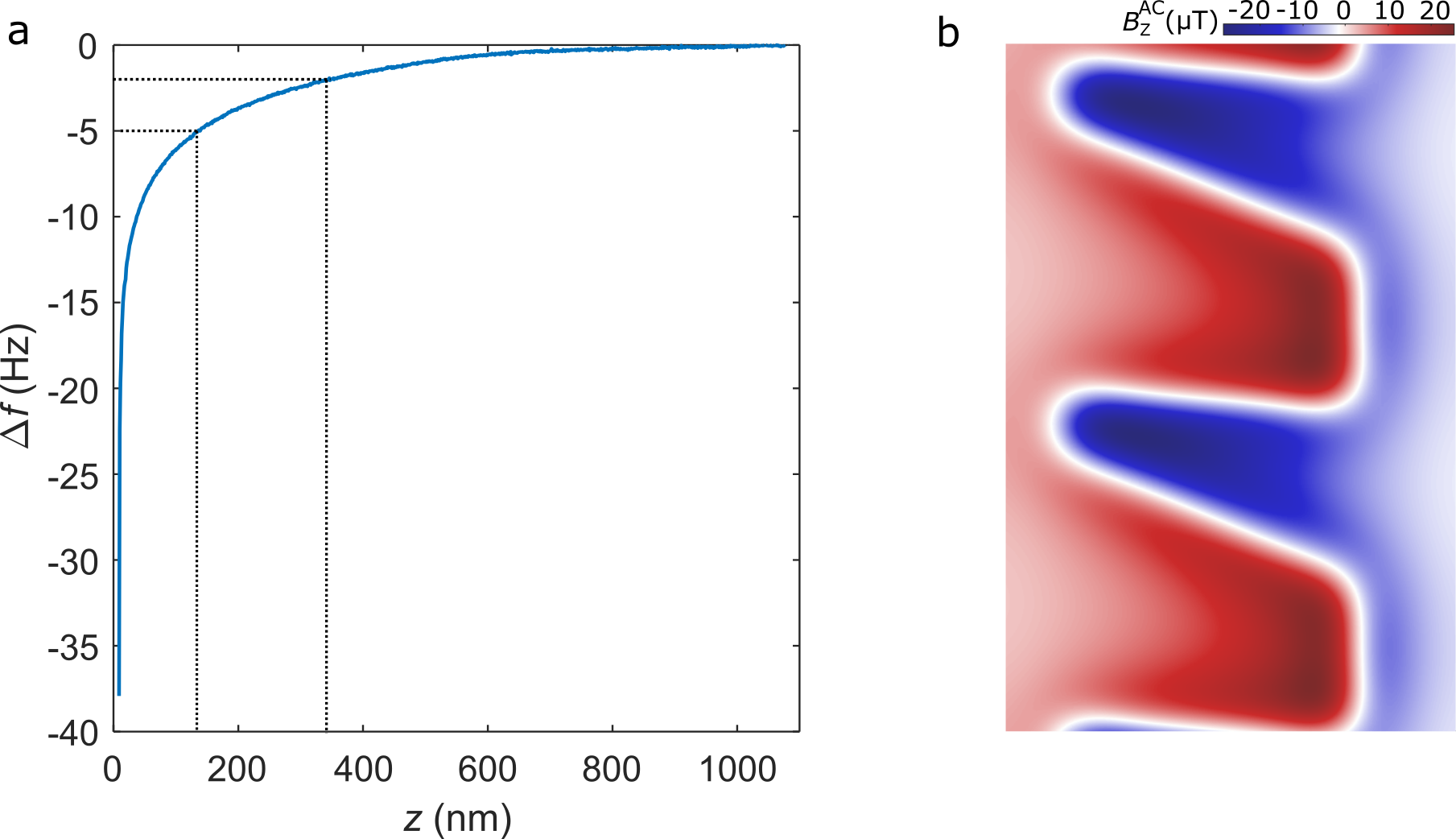}%
	\caption{ (a) Approach curve showing frequency shift $\Delta
          f$ as a function of tip-sample spacing $z$.  Dotted lines
          indicate the scanning height for the images of the
          current-carrying wire (Fig.~3 in the main text;
          $\Delta f = -5$~Hz) and images of the artificial spin ice
          (Fig.~4 in the main text; $\Delta f = -2$~Hz). (b)
          Simulation of the Biot-Savart field produced by the
          current-carrying wire at a SQUID-sample height of 520~nm.}%
	\label{S4}
\end{figure}

\subsection{Non-contact AFM}
Fig.~\ref{S4}~(a) shows an approach curve, i.e.\ a measurement of the
cantilever frequency shift $\Delta f$ as a function of the tip-sample
spacing $z$.  The sample being approached is the current-carrying wire
imaged in Fig.~2 of the main text.  The non-contact AFM image shown in
Fig.~3~(b) of the main text is carried out at a set-point
$\Delta f = -5$~Hz, which corresponds to a tip-sample spacing
$z = 140$~nm and a SQUID-sample spacing $z_{\text{SQUID}} = 315$~nm
(the height of the protruding tip is 175~nm).  The large tip-sample
spacing and the sample's interaction with the large plateau at the
apex of the cantilever yield the low spatial resolution apparent in
the AFM of Fig.~3~(b) of the main text. As we approach closer to the
sample, however, the interaction of the small protruding tip dominates
over that with the plateau. As a result, the non-contact AFM shown in
Fig.~4~(b) of the main text, which was taken at a tip-sample spacing
of $z = 65$~nm, is characterized by a resolution on the order of
50~nm.  A lower resolution `shadow' is visible in the image, due to
the contribution from the sample's interaction with the large
triangular plateau behind the protruding tip.

\subsection{Simulation of Biot-Savart field}
We simulate the Biot-Savart field produced by the current-carrying
wire. We use COMSOL Multiphysics 5.5 AC/DC module to simulate the
current density distribution inside the gold wire originating from an
applied current of 100~$\mu$A. The wire geometry matches the size of
the measured sample and we apply typical parameters for Au at
$T =4.2$~K. We then calculate the Biot-Savart field produced by the
current density and extract the $z$-component of the magnetic field
$B^{\text{ac}}_z$. For these simulations we use a custom tetrahedral
mesh with a minimum element size 80~nm. We filter the simulated
magnetic field at a SQUID-sample distance of 520~nm with a Gaussian
with standard deviation $\sigma = 365$~nm to account for the effective
diameter of the SQUID loop.  In Fig.~\ref{S4}~(b), we show the
simulation of $B^{\text{ac}}_z$ at the SQUID-sample spacing of the
measurement.  This simulation matches the measurement shown in
Fig.~3~(c) of the main text, confirming the scanning height obtained
from the approach curve shown in Fig.~\ref{S4}~(a).

\section{Mapping $dB_z/dz$ of the nanomagnet array}

\begin{figure}[h]
	\includegraphics[width=0.4\textwidth]{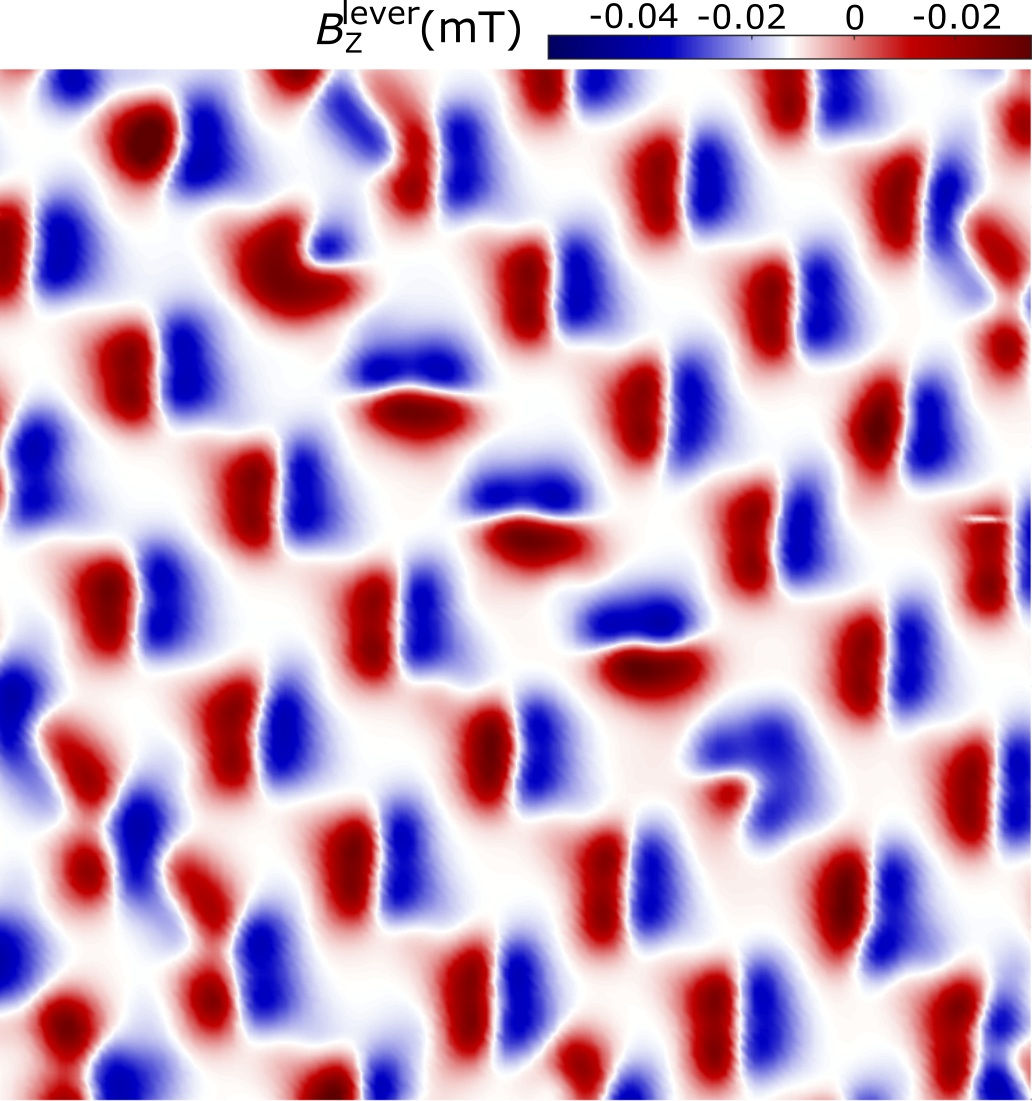}%
	\caption{Image of $z$-component of the magnetic field
          $B^{\text{lever}}_z$ at the cantilever oscillation frequency
          produced the artificial spin-ice sample.  This signal is
          proportional to the spatial derivative of $B_z$ along the
          cantilever oscillation direction $\hat{z}$. This image is
          analogous to that shown for the current-carrying wire in
          Fig.~3~(e) of the main text.}%
	\label{S5}
\end{figure}

The SQUID-on-cantilever is actuated at its fundamental mode ($f_0$ =
282 kHz) with an amplitude of 15~nm. By measuring the magneticl field
response of the SQUID at $f_0$, we obtain the spatial derivative of
the magnetic field along the cantilever oscillation direction
$dB_z/dz$. Imaging $dB_z/dz$ increases the sensitivity and spatial
resolution of the image compared to imaging magnetic
fields. Fig.~\ref{S5} presents $dB_z/dz$ measured above the square
lattice of nanomagnets in the type-II configuration with two type-III
vertices or monopoles.


%
\clearpage

%